# Mechanically stacked 1 nm thick carbon nanosheets: Ultrathin layered materials with tunable optical, chemical and electrical properties**

Christoph T. Nottbohm, Andrey Turchanin[*], André Beyer,

Rainer Stosch, Armin Gölzhäuser

[*] Dr. C.T. Nottbohm, PD Dr. A. Turchanin*, PD Dr. A. Beyer, Prof. A. Gölzhäuser

Fakultät für Physik, Universität Bielefeld, 33615 Bielefeld (Germany)

E-Mail: turchanin@physik.uni-bielefeld.de

Dr. R. Stosch

Physikalisch-Technische Bundesanstalt, 38116 Braunschweig (Germany)

[*] We thank Dr. B. Völkel and M. Büenfeld (University of Bielefeld) for technical assistance. Financial support from the Volkswagenstiftung, the Deutsche Forschungsgemeinschaft and BMBF is acknowledged.

**Abstract**

Carbon nanosheets are mechanically stable free-standing two-dimensional materials with a thickness of ~1 nm and well defined physical and chemical properties. They are made by radiation induced cross-linking of aromatic self-assembled monolayers. Here we present a route to the scalable fabrication of multilayer nanosheets with tunable electrical, optical and chemical properties on insulating substrates. Stacks up to five nanosheets with sizes of ~1 cm$^2$ on oxidized silicon were studied. Their optical characteristics were investigated by visual inspection, optical microscopy, UV/Vis reflection spectroscopy and model calculations. Their chemical composition was studied by X-ray photoelectron spectroscopy. The multilayer samples were then annealed in ultra high vacuum at various temperatures up to 1100 K. A subsequent investigation by Raman, X-ray photoelectron and UV/Vis reflection spectroscopy as well as by electrical four-point probe measurements demonstrates that the layered nanosheets transform into nanocrystalline graphene. This structural and chemical transformation is accompanied by changes in the optical properties and electrical conductivity and opens up a new path for the fabrication of ultrathin functional conductive coatings.

**Table of contents text and figure**

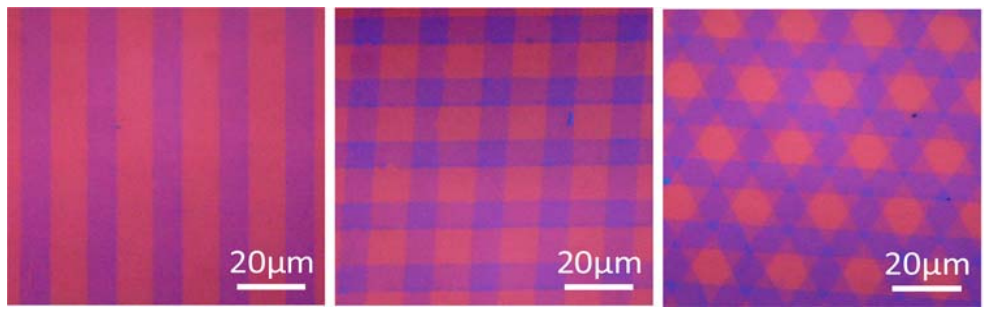

We present a route to the scalable fabrication of carbon nanosheet stacks on insulating substrates with a thickness of individual layers of ~1 nm and tunable electrical, optical and chemical properties.

# 1 Introduction

Free-standing two-dimensional (2D) nanomaterials, i. e. sheets or membranes with a thickness of only one or few molecules or atoms, recently have received a lot of attention.[1-6] They were proposed for novel applications such as ultrasensitive resonators and transducers in nanoelectromechanical systems (NEMS) as well as for functional coatings in biosensors or for nanoscale electronics.[5,6] Mechanically stable 2D nanomaterials also allow new pathways[3] for the controlled synthesis of novel composite materials, for example by the stacking of individual sheets in a layer-by-layer approach. Attempts to artificially assemble 2D layered materials date back to the 1930's when I. Langmuir and K. Blodgett[7] transferred molecular layers from a liquid surface onto a solid substrate. The Langmuir-Blodgett (LB) process is still widely used and has recently been extended to fabricate polymer LB-films.[8] In the 1990's G. Decher developed the layer-by-layer (LbL) assembly technique[9,10], where a surface is subsequently dipped into solutions of differently charged polyelectrolytes that then form stable multilayer assemblies via electrostatic interactions. However, both of these methods are liquid based and impose restrictions on the stacked molecules, such as an amphiphilic structure for LB or the presence of intramolecular charge for LbL. In addition, the individual LB layers are not mechanically stable, which limits their use. It is thus desirable to develop a universal method for the stacking of ultrathin 2D layers that works in ambient and does not depend on the specific interactions between the individual layers. Ideally each new layer can be mechanically placed onto the underlying one in a way that all layers still maintain their individual thickness, structure and chemical composition. If such a mechanical placing of 2D layers can be performed, this procedure would open an opportunity to build composite or gradient materials by a controlled stacking of functional layers.

In this contribution, we explore the mechanical stacking of ~1 nm thick free-standing sheets made from aromatic *self-assembled monolayers* (SAMs). SAMs are well ordered aggregates of organic molecules that spontaneously form into 2D molecular structures on a variety of solid surfaces.[11,12] SAMs are routinely used to modify the surfaces of different materials and are becoming key components for an *interfacial systems chemistry* that covers a variety of processes in fields as diverse as molecular electronics, electrochemistry, tribology and biotechnology.[13] It was recently shown that via electron or EUV radiation induced cross-linking[14-16] aromatic SAMs can be converted into mechanically stable 2D *carbon nanosheets* with a high thermal, chemical and mechanical stability[17-21]. The carbon nanosheets can be released from their original substrate and prepared as free-standing nanomembranes.[22,23] Moreover, these nanomembranes can be perforated with defined pore sizes[24] in the nanometer range as well as chemically functionalized[25], chemically patterned[26] or patterned with metal nanostructures[20,21]. It was also shown that during vacuum annealing the insulating nanosheets become conducting as they undergo a transformation to nanocrystalline graphene[23,27]. This combination of properties makes carbon nanosheets promising for the applications as individual building blocks to form ultrathin free-standing layered structures.

Here we explored mechanically stacked carbon nanosheets made from SAMs of 1,1'-biphenyl-4-thiol (BPT) on gold. We built stacks of up to five nanosheets on oxidized silicon and investigated their optical properties by visual inspection, optical microscopy, UV-Vis spectroscopy and model calculations. Their chemical composition was characterized by X-ray photoelectron spectroscopy (XPS). The multilayer samples were then annealed at various temperatures up to 1100K in ultra-high vacuum. The associated structural changes were investigated in ambient at room temperature (RT) by Raman spectroscopy whereas the changes in the optical properties, chemical composition and electrical conductivity were characterized by UV-Vis spectroscopy, XPS and four-point probe measurements, respectively.

## 2 Results

### 2.1 Preparation of multilayer stacks

Carbon nanosheets of cross-linked BPT SAMs are released from their gold substrate by a simple procedure[22,23], Scheme 1. First, poly(methyl methacrylate) (PMMA) is spun on top of the nanosheet. The ~500 nm thick PMMA film preserves the structural integrity of the 1 nm thick nanosheet so that gold substrate can be removed by wet-etching (see Experimental for details). The PMMA/nanosheet-sandwich is then transferred to a new substrate, e.g. silicon oxide, and the PMMA is finally dissolved in acetone, so that only the nanosheet remains on $SiO_2$. This procedure can be repeated until multiple layers of nanosheet are stacked on top of each other, Scheme 1. The mechanical transfer can be performed in a very controlled manner, as demonstrated in Fig.1, that shows an optical micrograph of ~10 µm wide nanosheet ribbons that were placed on a silicon substrate with a 300 nm thick layer of $SiO_2$.[1] The nanosheet ribbons were made by photolithography and plasma etching on the gold substrate before the transfer to $SiO_2$/Si. The ribbons appear darker and blue-shifted with respect to the substrate color. Figs. 1b, c show optical micrographs after additional layers of ribbons were placed on top of the first layer. It is clearly seen that two or three layers stacked on each other yield a higher contrast and blue-shift in the optical micrograph. The spatial orientation of the nanosheet ribbons can be well controlled during the manual transfer. In Fig. 1b an angle of 90° between two layers and in Fig. 1c angles of 60° between three subsequent nanosheet ribbon layers were chosen, the actual angles observed after the transfer deviate less than 3° from these values. For spectroscopic and microscopic characterization, samples of multilayer stacks with large homogeneous areas up to ~1 cm$^2$, as exemplary shown in Fig. 1d, were used.

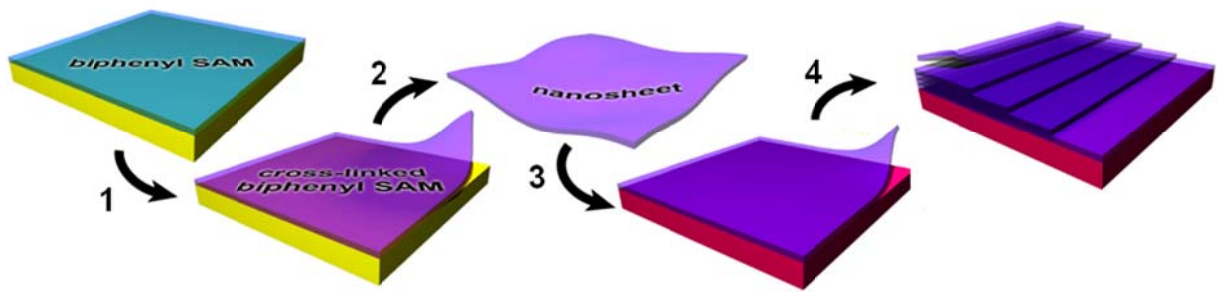

**Scheme 1:** Biphenyl-based self-assembled monolayers (SAMs) are crosslinked by electron irradiation (1) and subsequently removed from their substrate as nanosheets (2). These can then be positioned on arbitrary new substrates (3). By repeating this procedure (4) multilayer stacks of nanosheet can be produced.

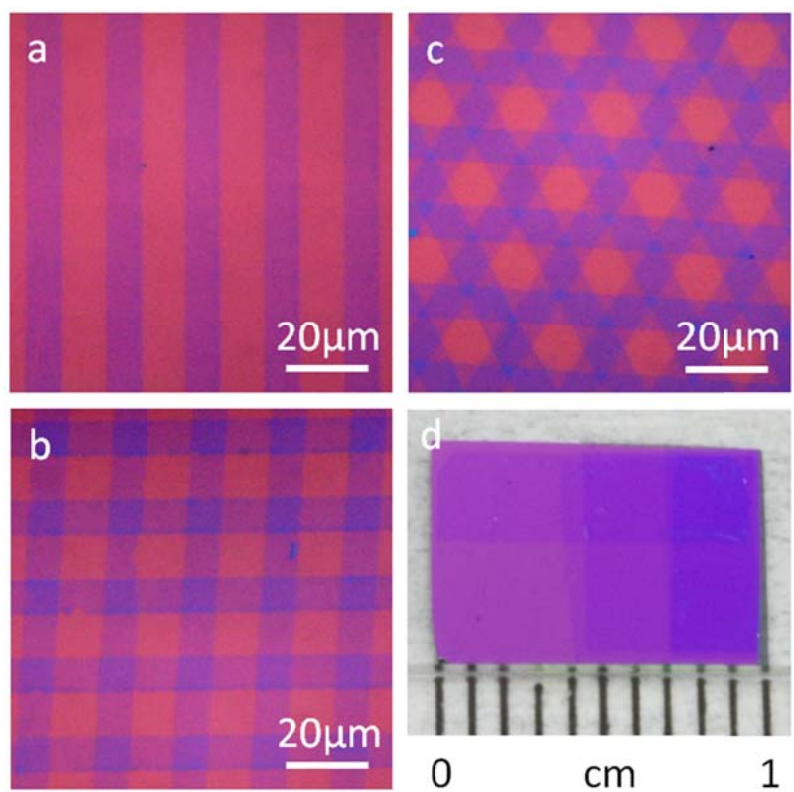

**Figure 1:** a) - c) Line patterns of nanosheet made by photolithography and plasma etching after transfer to $SiO_2$ substrates. The nanosheet appears blue shifted with respect to the substrate color. a) A single layer of 10 µm nanosheet ribbons. b) Two layers of nanosheet ribbons transferred at a ≈90° angle. c) Three layers at an angle of ≈60°. Each additional layer gives rise to a further blue shift. d) Stack of 1 to 5 nanosheets on $SiO_2$.

## 2.2  X-ray photoelectron spectroscopy

To determine the chemical composition and thickness the multilayer samples on SiO$_2$ were analyzed by XPS. We compare samples of one, three and five nanosheets that were annealed at 650 K, as for some non-annealed samples a small amount of residual PMMA from the transfer procedure was detected in the C1s signal (not shown). The presence of these contaminations depends strongly on the chain lengths of the PMMA and shorter chains yield "cleaner" nanosheets. The residual PMMA contaminations could be removed by annealing in vacuum at moderate temperatures (600-700 K) that do not introduce structural changes in nanosheets.[16,23] XPS results are summarized in Fig. 2. As seen from Fig. 2a, the C 1s signal demonstrates a main peak at 284.2 eV that is assigned to the aromatic carbon atoms in the cross-linked BPT and a weaker shoulder at 285.2 eV that is attributed to the carbon directly bonded to sulfur, as well as shake-up satellites at ~287 and ~290 eV[19]. Fig. 2b shows the S 2p region with two sulfur signals, each consisting of a spin-orbit split doublet with an intensity ratio of 1:2 and an energy separation of 1.2 eV. The doublet at 163.5/164.7 eV is assigned to disulfides or thiols [16,24] and the doublet at 168.3/169.5 eV is attributed to oxidized sulfur species[28] that were formed during or after transfer. Interestingly, the intensity of these oxidized sulfur species is roughly constant for one to five layers of nanosheet. If oxidation was equally present in every layer, the signal should increase with each additional layer.

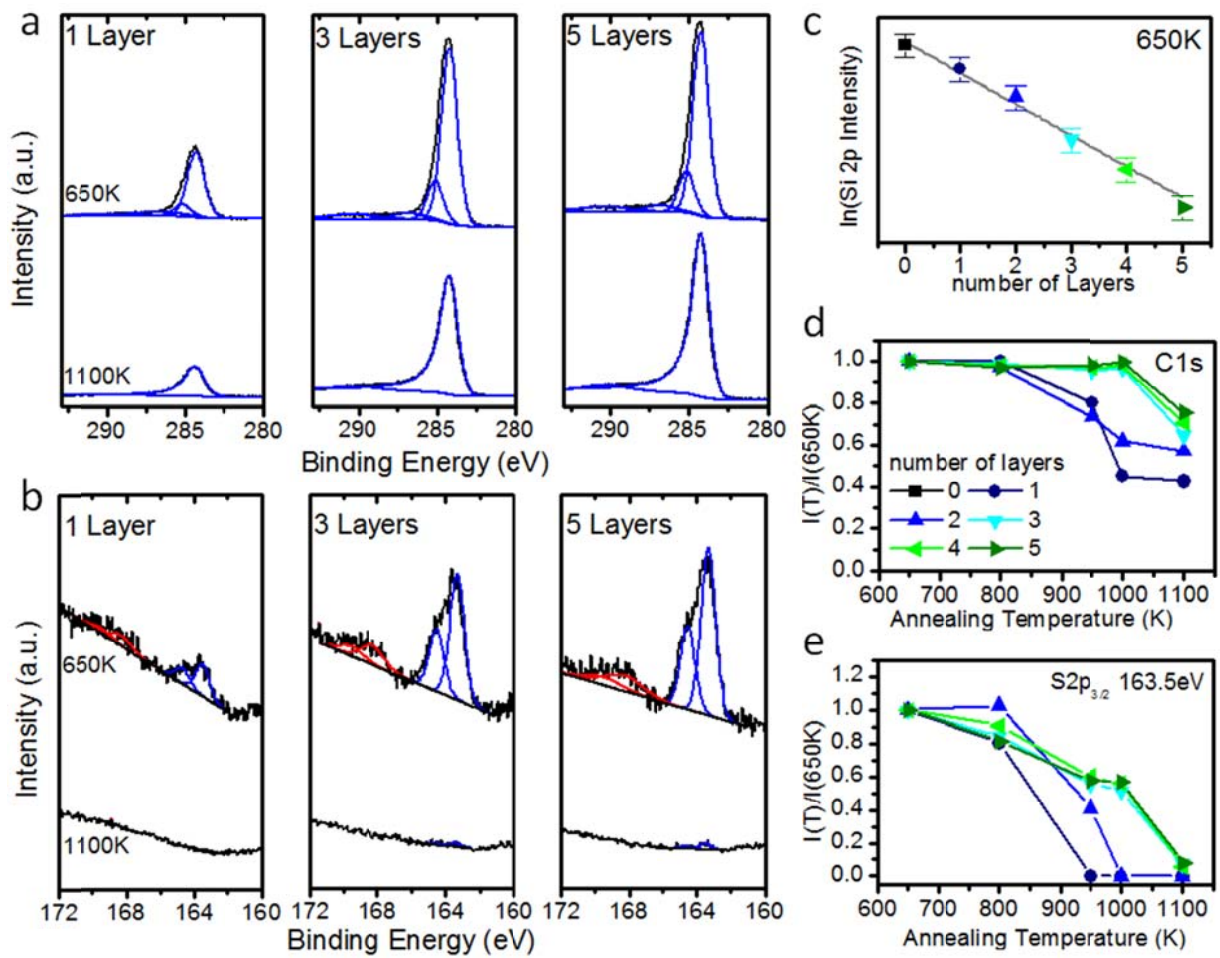

**Figure 2**: Summary of the XPS characterization of nanosheet multilayer stacks. a) C1s spectra of the clean nanosheet after transfer to $SiO_2$ for 1, 3 and 5 layers (top) as well as for samples annealed to 1100 K (bottom). The respective S2p spectra are shown in (b). c) shows the attenuation of the Si2p signal as a function of multilayer thickness. d) Intensity of the C1s signal for multilayer samples as a function of annealing temperature. The respective curves for both S2p signals are shown in (e).

The constant intensity is thus indicative that oxidation occurs during the handling of the nanosheets in air and mostly affects the sulfur close to the sample/air interface, that is in the topmost layer. Occasionally, a small iodine signal (not shown) was found that most likely remained from the etching and it was disappeared after annealing above 600 K. Finally, the effective thickness of the nanosheet layers was derived from the attenuation of the photoelectrons from the $SiO_2$ substrate (Si 2p, attenuation length λ=3.43 nm and O 1s, λ=2.67 nm)[29], see Fig. 2c. The average thickness per layer was found to be 0.85 ± 0.05 nm (0.77 ± 0.03 nm from O 1s), which is slightly lower than the ~1 nm, as expected from AFM [23].

It was recently found that nanosheet monolayers on gold substrates undergo a phase transition to nanocrystalline graphene phase upon annealing to 1100 K.[23] To study whether a similar transformation occurs in multilayer nanosheets on silicon oxide, we annealed them at temperatures up to 1100 K. The evolution of the C 1s intensity as a function of annealing temperature is displayed in Fig. 2d. For one and two layers the normalized intensity drops at temperatures above 800 K and levels off at 1100 K at ≈0.4 (one layer) and ≈0.6 (two layers), respectively. Thicker layers show a higher temperature resistance, the normalized intensity remains almost constant up to 1000 K and drops to ≈ 0.75 at 1100 K (for 5 layers). During annealing the shoulder at 285.2 eV disappears, whereas the main peak at 284.3 eV develops a pronounced asymmetry that can be described by a Doniach-Sunjic peak shape[30-32]. Such a peak shape is observed for the C 1s XP spectra (not shown) of freshly cleaved highly oriented pyrolytic graphite (HOPG).

The temperature behaviour of the sulfur signals is summarized in Figs. 2b and e. The oxidized sulfur species (S $2p_{3/2}$ 168.3 eV) disappears below 800 K, whereas the disulfides/thiols (S $2p_{3/2}$ 163.5 eV) show ~75% of the initial intensity. Further heating to 950 K removes all sulfur from the monolayer. For the two layer sample, the sulfur signal vanishes completely at 1000 K. For the three, four and five layer samples ~60% of the initial intensity is still observed at 1000 K. But after annealing at 1100 K, nearlly all sulfur is removed from those multilayer stacks. This results suggest that above 800 K a carbon-sulfur bond cleavage occurs in single nanosheets, however, the desorption of sulfur from the multilayers occurs at higher temperatures, possibly because thicker layers are more efficient as diffusion barriers.

The effective layer thickness of the multilayer samples decreases as a function of annealing temperature (supplementry Fig. 1). The values derived from the Si 2p and O 1s attenuation are in good agreement with each other and suggest a layer thickness of 0.55 ± 0.06 nm after annealing at 1100 K. Note that the attenuation length used in the calculation does not account for changes in the material properties, such as graphitization[23]. Using the attenuation length for graphite, λ(Si 2p)=2.37nm[33], the average layer thickness is calculated to ~0.35 nm.

### 2.3 Raman spectroscopy

After annealing structural transformations in the multilayer samples were characterized at RT by Raman spectroscopy with different excitation wavelengths (488, 514, 532 and 633 nm). The appearance of the so-called D- and G-peaks in the spectra, their positions, shapes, intensities as well as the wavelength dispersion are characteristic for the formation and degree of order in sp$^2$-hybridized carbon networks[34,35]. Fig. 3a shows Raman spectra of 1 to 5 nanosheet layers on silicon oxide measured with 488 nm excitation wavelength after annealing at 1100 K. The graphitization is confirmed by the presence of both D- and G-peaks in the spectra. In the non-annealed samples these peaks are not present[23]. The intensity of the D- and G-peaks increases with the number of layers in a stack. Their position and the I(D)/I(G)-ratio correlate with the number of layers and excitation wavelength (Figs.3 b, c). For all excitation wavelengths, the position of the G-peak shifts towards lower wavenumbers as the stack increases from 1 to 5 layers. This shift increases from ~5 cm$^{-1}$ for 488 nm excitation to ~7 cm$^{-1}$ for 633 nm excitation. On the contrary, the D-peak shifts towards higher wavenumbers with an increasing number of layers, this increase is ~12 cm$^{-1}$ for 488 nm excitation and ~4 cm$^{-1}$ for 633 nm excitation. Moreover, wavelength dispersion of the D-peak was observed for individual stacks. Its wavenumber decreases by ~14 cm$^{-1}$ and ~24 cm$^{-1}$ for 1 and 5 layers, respectively, with the increase of the excitation wavelength from 488 to 633 nm. As seen from Fig. 3b, the I(D)/I(G)-ratio also depends on the stack thickness and excitation wavelength. Its magnitude decreases as the number of layers growths and becomes less pronounced towards longer wavelengths. All observations in the Raman spectra are in accordance with the formation of a network of nanocrystalline graphene[23,35] in the layered samples after annealing. Their crystallinity increases with the number of stacked nanosheets.

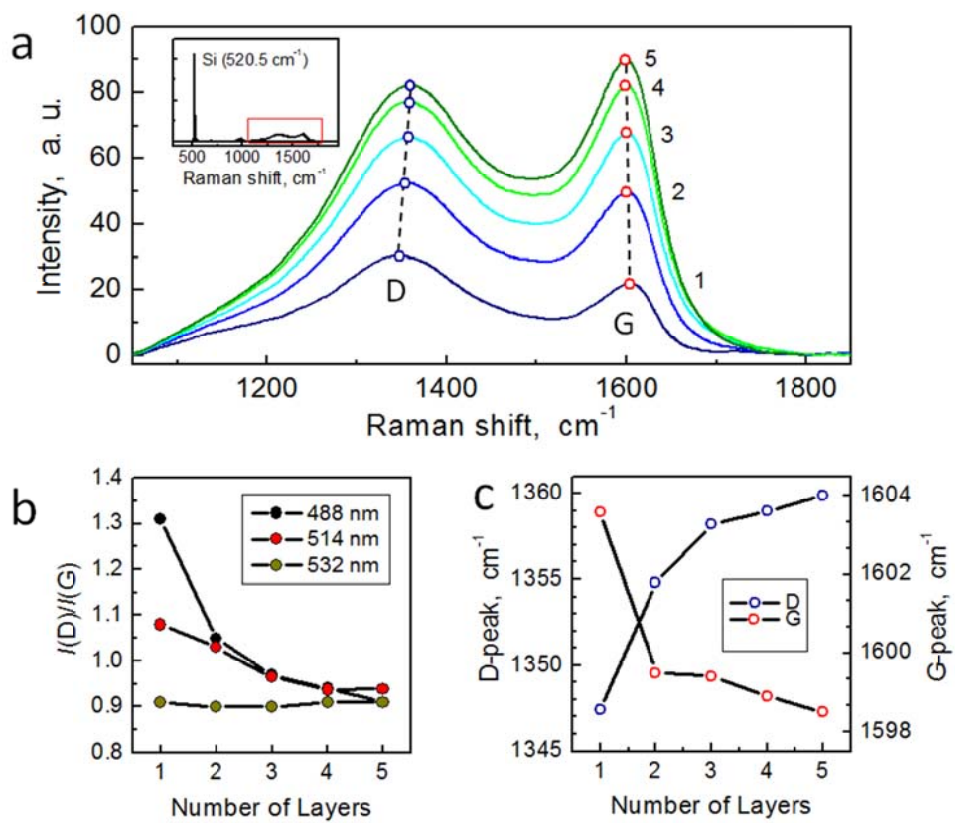

**Figure 3:** a) Raman spectra of 1 to 5 nanosheet layers after annealing at 1100 K taken at 488 nm excitation; b) D/G peak intensity ratio vs. number of layers for three excitation wavelengths; c) D- and G-peak dispersion vs. number of layers at 488 nm excitation.

## 2.4 Electrical measurements

The electrical conductivity of the nanosheet stacks is expected to change with the graphitization. Thus, for each annealing step the RT sheet resistivity ($\rho_\square$) of the annealed stacks was determined by four-point probe measurements, Fig. 4. All samples showed linear current-voltage curves, a typical example is plotted in Fig. 4b. The sheet resistivity as a function of annealing temperature is plotted in Fig. 4a. Non-annealed nanosheets are insulating and the onset of conductivity occurred after annealing at 800 K, where a sheet resistivity of 18 GΩ/sq was measured for a four layer stack. In a five layer stack, the sheet resistivity was one order of magnitude (2.5±1.3 GΩ/sq) smaller. After annealing at 950 K, $\rho_\square$ dropped two orders of magnitude to 21±6 MΩ/sq for five and 76±16 MΩ/sq for four layers. Three and two layer stacks were also conductive with sheet resistivities of 1.9±0.8 GΩ/sq and 14±10 GΩ/sq, respectively. Upon annealing to 1100 K the sheet resistivity dropped further to 10.8±2.9 kΩ/sq for five, 30.3±1.4 kΩ/sq for four, and 37.2±6.8 kΩ/sq, for three layers. The two layer sample had a $\rho_\square$ of 0.38±0.13 MΩ/sq and the single layer sample was still insulating at this temperature. These results correlate very well with the Raman spectroscopy measurements of the previous section. Note, as shown in our previous study[23], single nanosheets annealed on gold substrates at 1100 K are conducting at RT and have $\rho_\square$~$10^5$ Ω/sq. This difference in comparison to the results for silicon oxide substrate may result from the specific interactions at the silicon oxide /nanosheet interface, which postpone the graphitization.

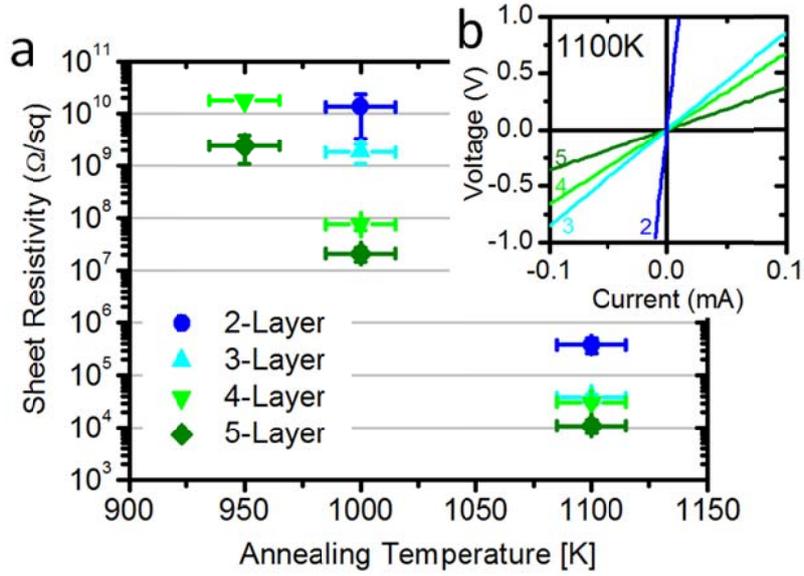

**Figure 4:** a) RT sheet resistivity of nanosheet multilayers as a function of annealing temperature. The error bars represent the standard deviation between measurements at different spots on the samples. b) RT linear current voltage curves for 2, 3, 4 and 5 layer samples after annealing at 1100 K.

## 2.5 UV-Vis reflection spectroscopy

Due to the interference contrast, it is possible to see a single layer of nanosheet by the naked eye if it is placed on top of an oxidized silicon wafer[23]. The origin of this contrast has been described for single and multilayer graphene by Casiraghi[36], Blake[37] and others.[38-43] It dependents mostly on the thickness of the oxide layer, that acts as a spacer similarly to an interferometer. We have found that the contrast is greatly enhanced after annealing, indicating a change of optical properties during graphitization. To understand the origin of this contrast for nanosheets, UV/Vis spectra were measured with a reflection probe. The reflectance $R_\lambda$ is given by

$$R_\lambda = \frac{I(sample)}{I(substrate)} \qquad (1)$$

where *I(sample)* and *I(substrate)* are the intensities of the reflected light from the silicon oxide/silicon substrate with and without nanosheet, respectively. The contrast δ is defined as the difference between the substrate and sample intensities, normalized to the substrate intensity, so that

$$\delta = 1 - R_\lambda \qquad (2)$$

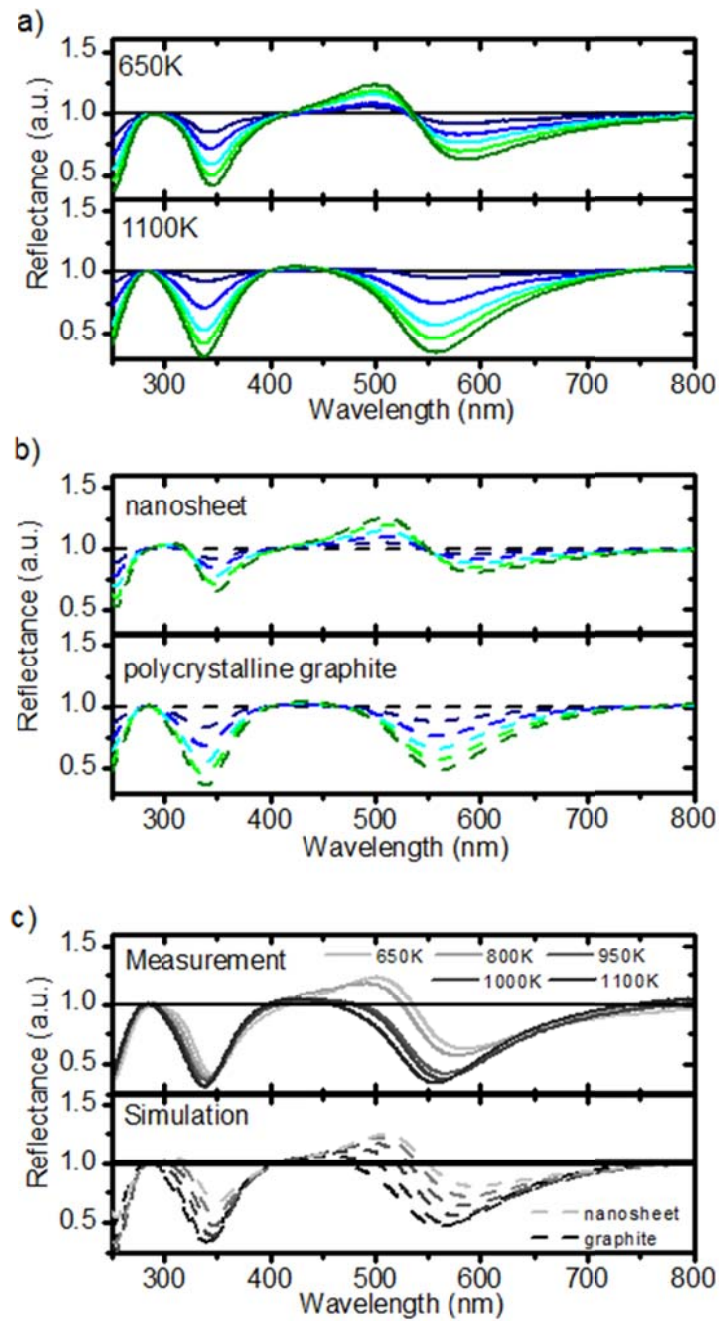

**Figure 5**: a) UV/Vis reflection spectra of nanosheet multilayers on SiO$_2$/Si (top) and after annealing to 1100 K (bottom). b) Simulation of the spectra from (a). c) Reflection spectra of the 5 Layer sample as a function of annealing temperature (top) together with simulations that assume a homogeneous mixture of the nanosheet with graphite in steps of 20%.

Regions darker than the substrate have a positive sign of the contrast and vice versa.

Fig. 5a shows the measured spectra of the nanosheet (top). The spectra have pronounced minima at λ≈250, 350 and 570 nm as well as a maximum at λ≈500 nm. With each additional layer, the reflected intensity at the maxima (minima) is roughly linearly increasing (decreasing) and the position is slightly shifted towards higher wavelengths. After annealing to 1100 K (bottom spectra, Fig. 5a), the maximum at λ≈500 nm disappears. The magnitude of the minima increases and shifts to lower wavelengths. A linear decrease of the reflected intensity was observed for each additional layer. Compared to the spectra before annealing, the position of the minima is approximately constant.

## 3  Discussion

The fabrication of ultrathin 2D films with homogenous thickness and well defined physical and chemical properties is a challenging task with applications in physics, chemistry and materials science.[44] By mechanical stacking of ~1 nm thick carbon nanosheets from aromatic SAMs and the thermal treatment of these multilayers we have shown a promising path towards the production of such materials. Multilayer stacks of nanosheets were structured with high accuracy and over large areas, Fig.1. Their total thickness is simply given by the number of layers, which is apparent from the attenuation of the substrate signal in XPS (cf. Fig. 2c and supplementary Fig. 1). Moreover, each layer possesses its individual chemical functionality[24,25], which opens up new opportunities to create composite materials by the incorporation of nanomaterials with different functionalities into the stacks. On the one hand, metallic nanostructures could be immobilized at the surface of the individual nanosheets[20-22], on the other hand organic molecules (e.g. fluorophores or proteins[25,45,46]). It seems feasible to stack alternating conductive and insulating layers to fabricate small capacitors.

The homogeneity of the nanosheet multilayer stacks has been confirmed also from the optical reflection spectra in Fig. 5a. Each additional layer gives rise approximately to a linear increase of contrast. This behavior can be understood by comparing the reflection data with simulations. As the refractive index of the BPT nanosheet has not yet been determined experimentally, for the simulations it was estimated from spectroscopic data (Fig. 5a). Polymeric materials containing aromatic units generally have real parts of the refractive indices between 1.5 and 1.7, so for the simulations n=1.7 was used as this is close to literature values for polyphenylenes[47]. This value was kept constant for all wavelengths. To account for absorption in the range below 500 nm, an imaginary part k was added to the refractive index which is based on the absorption of aromatic model compounds[48-50]. For the calculations a nanosheet thickness of 1 nm was used. The calculated spectra are presented in Fig. 5b and are in good qualitative agreement over the whole spectral range with the experimental data in Fig. 5a. The absolute values slightly deviate from the experimental data, possibly because of the exact values of the refractive index are unknown. For a more accurate description the complex refractive index of the BPT nanosheets has to be determined, for example by spectroscopic ellipsometry.

The chosen complex refractive index of the nanosheets was then used to simulate their optical contrast on the oxidized silicon. In Fig 6a,c the contrast δ is plotted against the $SiO_2$ thickness, the wavelength and the number of layers. Fig. 6a shows the contrast for one layer of BPT nanosheet. It has a ray-like appearance due to a shift of the maxima and minima of the substrate reflectance ($R_\lambda$=I($SiO_2$/Si)/I(Si)) towards higher wavelengths for higher oxide thicknesses, which is given by the criteria for constructive and destructive interference in thin films (supplementary Fig. 2). The maximum contrast for all substrate thicknesses is found in the UV-range, which is enhanced due to absorption. For $SiO_2$ thicknesses of 70 nm or more, both positive and negative contrast is found. The strong wavelength dependence of the contrast requires carefully selecting the thickness of the $SiO_2$ layer to maximize the visibility of nanosheet multilayers. Also the contrast can be improved by the use of narrow band filters. In Fig. 6c the contrast is plotted as a function of nanosheet layers for a fixed $SiO_2$ substrate thickness of 278.5 nm. As expected for interference in thin films, with each additional layer the maxima and minima are shifting to longer wavelengths, accompanied by a steady contrast enhansment.

As can be seen from XPS, Raman spectroscopy, optical and electrical measurements, after annealing at 1100 K, the nanosheets undergo a transition towards nanocrystalline graphene. An onset of the graphitization was found at ≈800 K, where a RT sheet resistivity, $\rho_\square$, of about $10^9$ and $10^{10}$ Ω/sq was detected for five and four layer samples, respectively (cf. Fig. 4). By annealing to 1100 K, the graphitization of the samples continues and $\rho_\square$ drops to ~$10^4$ Ω/sq for the five layer sample. As was demonstrated by Raman spectroscopy, the crystallinity of the nanosheet stacks increases with the number of layers. However, a single nanosheet on $SiO_2$ annealed at 1100K is still not conducting. Hence, multilayers provide an opportunity to flexibly tune the conductivity of 2D carbon nanosheet systems by either controlling the annealing temperature or the number of layers in a stack. This is also reflected in the optical characteristics.

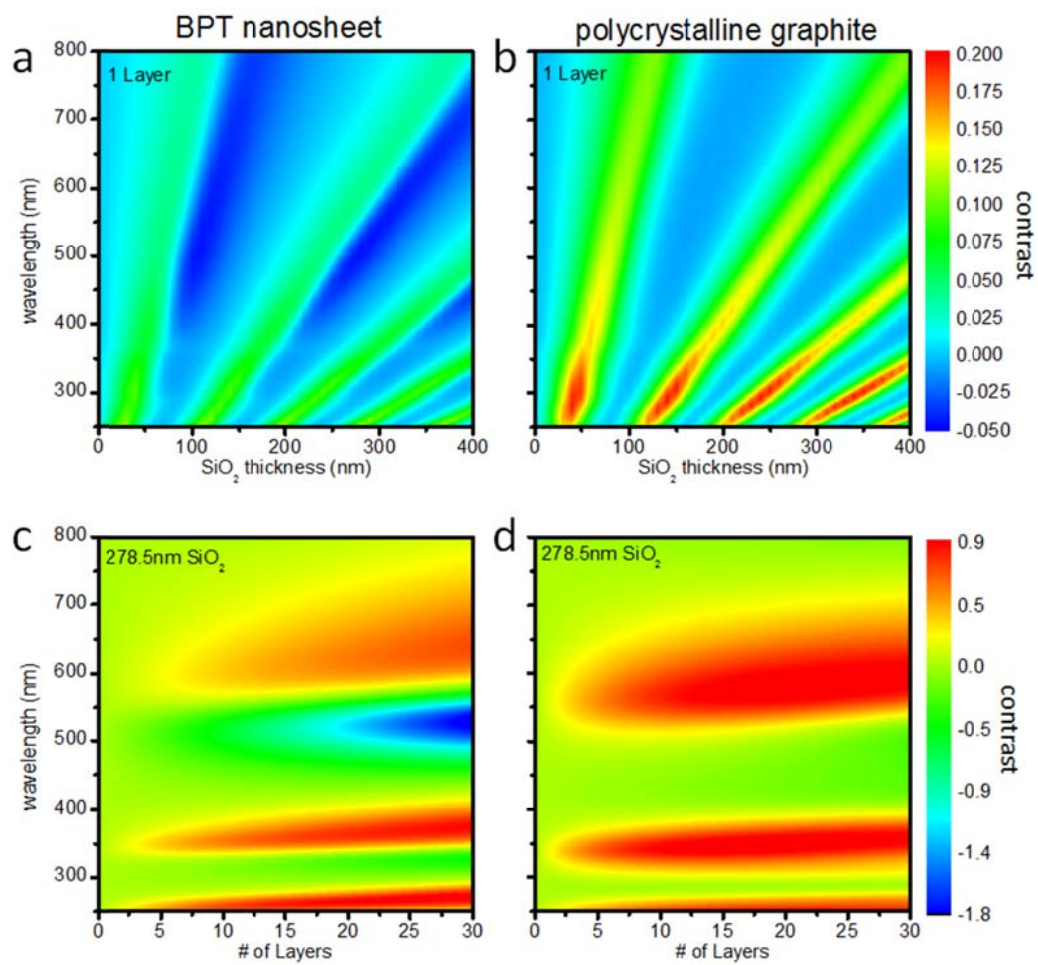

**Fig. 6:** Simulation of the contrast of one layer of nanosheet (a) and polycrystalline graphite (b) as a function of wavelength and thickness of the $SiO_2$ substrate. c, d) Contrast as a function of wavelength and layer thickness of nanosheet and graphite at a constant substrate $SiO_2$ thickness of 278.5 nm.

By comparing the reflection spectra of non-graphitized and graphitized nanosheets (cf. Fig. 5a), it is apparent that the graphitization gives rise to different spectral features in the UV-Vis range. The most prominent difference is the maximum in reflectance around 500 nm. In pyrolized stacks such feature is not observed. This can be understood by simulating reflection spectra for thin graphitic layers. To this end, the refractive index for polycrystalline graphite[51] was used. The thickness of the individual layers was used as determined by XPS (0.35 nm). Fig. 5b (bottom spectra) shows the calculated reflectance data of the graphitized nanosheets. As can be seen, the simulation corresponds well to the experiment (cf. Fig. 5a). However, the absolute values of the calculated reflectance show somewhat higher values in comparison to the experiment. This behavior is most pronounced in the range from 500 to 600 nm and may be indicative for preferentially in plane oriented graphene patches in the pyrolized stacks. We have found that polycrystalline graphite and oriented graphene give rise to similar spectral features in the UV-Vis. However, due to the anisotropy of the refractive index of graphene, the contrast is higher than that of polycrystalline graphite of the same thickness at normal incidence.

In Figs. 6b and d the calculated contrast of thin polycrystalline graphite layers is presented together with the data for BPT nanosheets (Figs. 6a and c). Due to its strong absorption, graphite (Fig 6b) shows a higher contrast than the nanosheet (Fig.

6a) over the whole spectral range and all substrate thicknesses. The highest maximum is found in the UV range, which can be attributed to the π → π* transition of graphite. [36,52] Figs. 6c and d show the contrast of nanosheet and graphite as a function of wavelength and the number of layers at the given $SiO_2$ thickness of 278.5 nm. For thicker layers the contrast for graphite saturates in the maxima at about 0.9, whereas for the nanosheet it still grows in the minima. For multilayer stacks with a thickness of several tens of nanometers, however, special substrates for enhancing the visibility are no longer necessary. These findings are in accordance with those published for the contrast of graphene on oxidized silicon substrates.[36,37]

The gradual graphitization of nanosheets upon annealing is reflected not only in the electrical (Fig. 4) but also in the optical properties, as shown for the five layers sample in Fig. 5c. The changes in the UV-Vis reflection spectra during this transformation can be represented by assuming the formation of a homogenous mixture of graphene in a polymeric matrix. In this case the refractive index of the mixture is calculated from the refractive indices of both components. The most commonly used mixing rule, the so called Lorentz-Lorenz formula [53-55], was utilized for calculations. Simulated spectra for the transition from nanosheet to nanocrystalline graphene in steps of 20% are shown in Fig. 5c (bottom). The decrease of the average layer thickness, as detected by XPS, was calculated assuming a linear dependence. As can be seen from Fig. 5c, the suggested model describes the experiment well and can be used to estimate the degree of graphitization after annealing from the optical data.

# 4 Conclusion

We have introduced a route to scalable and controllable fabrication of ultrathin layered 2D carbon materials by simple mechanical stacking of free-standing ~1 nm thick nanosheets from electron cross-linked aromatic self-assembled monolayers (SAMs). As shown by complimentary spectroscopic techniques (XPS, Raman and UV-Vis spectroscopy), optical microscopy and electrical measurements their homogeneous thickness, chemical composition, structural, electrical and optical properties can be flexibly tuned by the number of layers in the stacks and annealing temperature. Upon annealing in vacuum, the nanosheet multilayer undergoes a transition to a nanocrystalline graphene phase, which is reflected in both electrical conductivity and optical properties. The optical contrast of the non-annealed and annealed mono-/multilayer of nanosheets on Si/$SiO_2$ substrate has been described on the basis of spectroscopic data and model calculations. The suggested route to scalably fabricate ultrathin 2D carbon layers with adjustable thickness, electrical and optical properties opens broad avenues to the engineering of novel functional 2D carbon-based materials for application ranging from electronics to biophysics.

## 5 Experimental

### 5.1 Sample preparation

#### 5.1.1 SAM preparation

For the preparation of 1,1'-biphenyl-4-thiol (BPT), 300 nm thermally evaporated Au on mica substrates (Georg Albert PVD-Coatings) were used. The substrates were cleaned in a UV/ozone-cleaner (UVOH 150 LAB from FHR, Germany) for 3-5 minutes at an oxygen flux of 1 l/min, rinsed with water and ethanol and blown dry in a stream of nitrogen. They were then immersed in a ≈1 mmol solution of BPT in dry, degassed dimethylformamide (DMF) for 72 h in a sealed flask under nitrogen. Afterwards samples were rinsed with DMF and ethanol and blown dry with nitrogen. Cross-linking was achieved in high vacuum (<$5*10^{-7}$ mbar) with an electron floodgun (Specs) at an electron energy of 100 eV and typical dose of 50 mC/cm$^2$.

#### 5.1.2 Transfer

The nanosheets were transferred to SiO$_2$/Si substrates by a procedure that was described previously in Ref. [22,23]. First, a transfer layer is applied to the sample. To this end, PMMA (AR-P 671.04) is spincast onto the nanosheet for 30 s at 4000 rpm as a transfer medium and cured on a hotplate for 5 min at 90°C. The Au/nanosheet/resist sandwich cleaved from the mica substrate by immersion in 48 % HF for 15-60 min, followed by carefully dipping it into water, so that the Au/nanosheet/resist sandwich is swimming on the water surface. By using a piece of silicon it can then be transferred to a KI/I$_2$ etching solution (0.6 mol/l KI, 0.09 mol/l I$_2$) where the Au is completely removed after 15 min[56]. The nanosheet/resist sandwich is then transferred back to water to prevent contamination with I$_2$. The monolayer is then swimming on the water surface and can be transferred by fishing it out with a new substrate. It is then carefully dried in a gentle flow of nitrogen. Best contact with the new substrate can be achieved by drying it on a hotplate at 90°C for several minutes. The resist is removed by immersion of the sample in acetone under light stirring for at least 2 minutes. Afterwards the sample is dipped in methanol and blown dry with nitrogen.

### *5.1.3 Patterning*

Patterned nanosheets were prepared by photolithography. To this end photoresist (AR-P 3510, Allresist) was spincoated onto the sample (30s@4000 rpm) and baked on a hot plate at 90°C for 2 min. Patterning was achieved by exposure through a chromium mask (delta mask) for 1.3 s in a mask aligner (Süss MJB 3). After development (AR-P 671.04, 60 s) the samples were rinsed with water and blown dry in a steam of nitrogen. Removal of the nanosheet in exposed regions was done by an oxygen plasma (see [57] for details on the homebuilt system) for 1 min. Afterwards the photoresist was removed with acetone and water under sonication for 5 minutes each. In some experiments also photoresist stripper AZ 400T (Clariant) was used.

### *5.1.4 Annealing*

Annealing of nanosheet multilayers on $SiO_2$ substrates was conducted in UHV conditions in Mo sample holders with a resistive heater with the typical heating/cooling rates of ~150 K/h and the annealing time of 0.5 h. Annealing temperature was controlled with a Ni/Ni-Cr thermocouple and two-color pyrometer (SensorTherm).

## 5.2 Spectroscopy and microscopy

XP-spectra were recorded with an Omicron Multiprobe spectrometer using monochromatic Al Kα radiation. Binding energies were calibrated with respect to the C 1s signal of the nanosheet at 284.2 eV. For peak fitting a Shirley background and a Gaussian-Lorentzian line shape with 30% Lorentzian were used.

Optical microscopy was performed at an Olympus BX51 equipped with a C5060 digital camera.

Raman spectra were collected using two micro Raman spectrometer (Horiba Jobin Yvon) operated in backscattering mode. Measurements at 488 and 514.5 nm were recorded on a T64000 triple-grating instrument equipped with an $Ar^+$-Laser, a 80x objective and a lq-$N_2$ cooled CCD detector (~ 1.5 $cm^{-1}$ spectral resolution). Spectra at 532 and 633 nm were obtained using a LabRam ARAMIS equipped with a frequency-doubled Nd:YAG-Laser and HeNe Laser, a 100x objective and a thermoelectrically cooled CCD detector ( 2-3 $cm^{-1}$ spectral resolution). The Si-peak at 520.5 $cm^{-1}$ was used for peak shift calibration of the instruments.

UV-Vis reflection spectroscopy was performed with an Ocean Optics USB2000 spectrometer using a reflection probe (Optocon) that was mounted perpendicular to the sample surface at a constant distance.

## 5.3  Simulation of reflection spectra

In order to simulate Reflection spectra for multilayer systems multiple reflections (from each layer) have to be considered. A generalized method for the analysis the optical properties of multilayer systems is the so called matrix method [53,58-60]. It considers the total electric and magnetic fields, as described by the Fresnel equations, in every layer. The boundary conditions are applied at every interface. The matrix method is implemented in a variety of commercial programs for simulation and optimization of antireflective coatings and filters. In this work the open source program OpenFilters by Larouche and Martinu has been used to simulate optical reflection spectra [61].

As a first step, the spectrum of the SiO$_2$ substrate was measured using a freshly etched Si wafer as a reference and matched with simulations to determine the precise SiO$_2$ thickness. In order to accurately simulate the spectral data, it is necessary to use the wavelength dependent complex refractive indices of both Si and SiO$_2$.[62] A thickness of 278.5 nm was found to best describe the measured data, and was used for all further simulations (see supplementary Fig. 2).

The acceptance angle at that light reflected from the sample can still enter the probe is dependent on the numerical aperture and the probe-sample distance. To account for all possible acceptance angles, all spectra were weighed by a Gaussian distribution.

### 5.4 4-Probe measurements:

Electrical measurements were done by a standard four probe setup using Suess probes and a Keithley source measure unit. The probes were equidistantly spaced on the sample. If the distance between the probes is small compared to the lateral dimensions of the sample, the sheet resistivity $\rho_\square$ for a thin film can be calculated by [63]

$$\rho_\square = \left(\frac{\Omega}{sq}\right) = \frac{\rho}{t} = \frac{\pi}{\ln(2)}\frac{U}{I} = 4.53\frac{U}{I}$$

where the resistivity ρ, the thickness of the thin film t, the Voltage U, and the current I. The measurement limit of sheet resistivity was ~100 GΩ/sq.

# *Supplementary information*

# Mechanically stacked 1 nm thick carbon nanosheets: Ultrathin layered materials with tunable optical, chemical and electrical properties**

Christoph T. Nottbohm, Andrey Turchanin[*], André Beyer,

Rainer Stosch, Armin Gölzhäuser

[*] Dr. C.T. Nottbohm, PD Dr. A. Turchanin*, PD Dr. A. Beyer, Prof. A. Gölzhäuser

Fakultät für Physik, Universität Bielefeld, 33615 Bielefeld (Germany)

E-Mail: turchanin@physik.uni-bielefeld.de

Dr. R. Stosch

Physikalisch-Technische Bundesanstalt, 38116 Braunschweig (Germany)

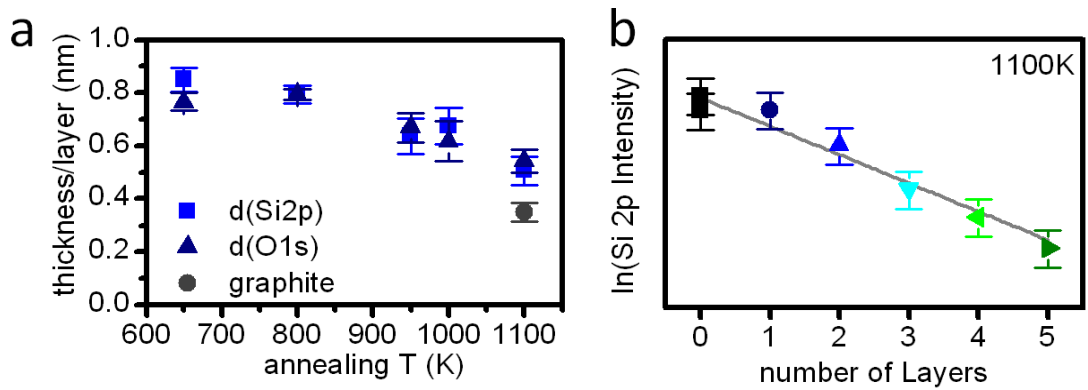

**Suppl. Fig. 1:** a) Average layer thickness of the nanosheets as a function of the annealing temperature as calculated from the attenuation of the Si 2p and O 1s XPS signals of the substrate. b) Plot of the Si 2p XPS intensity of the $SiO_2$ substrate as a function of nanosheet layers for samples annealed at 1100K.

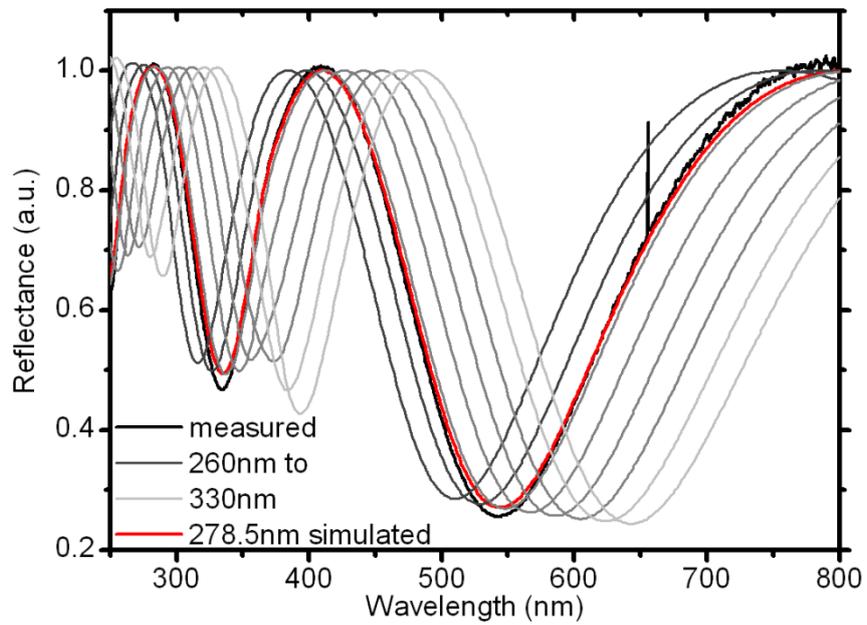

**Suppl. Fig. 2:** UV-Vis reflectance data of the $SiO_2$ substrate together with simulated spectra for substrates with different oxide thicknesses. A thickness of 278.5nm best describes the experimental data and was used for all model calculations.